\documentclass{article}%
\usepackage{amsfonts}
\usepackage{amssymb}
\usepackage{amsmath}
\usepackage{graphicx}%
\providecommand{\U}[1]{\protect\rule{.1in}{.1in}}
\begin{document}

\title{The new partitional approach to (literally) interpreting quantum mechanics}
\author{David Ellerman\\University of Ljubljana, Slovenia}
\maketitle

\begin{abstract}
\noindent This paper presents a new `partitional' approach to understanding or
interpreting standard quantum mechanics (QM). The thesis is that the
mathematics (not the physics) of QM is the Hilbert space version of the math
of partitions on a set and, conversely, the math of partitions is a
skeletonized set level version of the math of QM. Since at the set level,
partitions are the mathematical tool to represent distinctions and
indistinctions (or definiteness and indefiniteness), this approach shows how
to interpret the key non-classical QM notion of superposition in terms of
(objective) indefiniteness between definite alternatives (as opposed to seeing
it as the sum of `waves'). Hence this partitional approach substantiates what
might be called the Objective Indefiniteness Interpretation or what Abner
Shimony called the Literal Interpretation of QM.

\end{abstract}

\section{Introduction: The basic thesis}

The purpose of this paper is to expound a new way to interpret quantum
mechanics (QM), or, to be more precise, to interpret the mathematics (not the
physics\footnote{The physics of QM is obtained by the quantization of
classical physics. Our focus is on the specific nature of the
\textit{mathematical} framework of standard von Neumann/Dirac QM.}) of QM. The
key mathematical, indeed logical, concept is the notion of a partition on a
set--or equivalently, the notion of a quotient set or equivalence relation.
The basic thesis is that the math of QM is the Hilbert space version of the
math of partitions. Partitions are the basic logical concept to describe
distinctions versus indistinctions, definiteness versus indefiniteness,
distinguishability versus indistinguishability, or difference versus identity.
The key non-classical notion in QM is that of \textit{superposition}--with
entanglement being a particularly unintuitive special case. The result of the
thesis is to give a partitional explication of superposition in terms of
objective indefiniteness between definite alternatives--so that this approach
to QM could be called the \textit{Objective Indefiniteness or Literal
Interpretation} of QM.

\begin{quotation}
From these two basic ideas alone -- indefiniteness and the superposition
principle -- it should be clear already that quantum mechanics conflicts
sharply with common sense. If the quantum state of a system is a complete
description of the system, then a quantity that has an indefinite value in
that quantum state is objectively indefinite; its value is not merely unknown
by the scientist who seeks to describe the system. Furthermore, since the
outcome of a measurement of an objectively indefinite quantity is not
determined by the quantum state, and yet the quantum state is the complete
bearer of information about the system, the outcome is strictly a matter of
objective chance -- not just a matter of chance in the sense of
unpredictability by the scientist. Finally, the probability of each possible
outcome of the measurement is an objective probability. Classical physics did
not conflict with common sense in these fundamental ways. \cite[p.
47]{shim:reality}
\end{quotation}

Abner Shimony suggested calling this interpretation of the math or "formalism
of quantum mechanics" as the Literal Interpretation.

\begin{quotation}
\noindent These statements ... may collectively be called "the Literal
Interpretation" of quantum mechanics. This is the interpretation resulting
from taking the formalism of quantum mechanics literally, as giving a
representation of physical properties themselves, rather than of human
knowledge of them, and by taking this representation to be complete. \cite[pp.
6-7]{shim:vienna}
\end{quotation}

\section{The lattice of partitions}

A \textit{partition} $\pi$ on a set $U=\left\{  u_{1},...,u_{n}\right\}  $ is
a set of nonempty subsets or blocks $\pi=\left\{  B_{1},...,B_{m}\right\}  $
that are mutually disjoint and jointly exhaustive (their union is
$U$).\footnote{We stick to the finite case since our purpose is conceptual
rather than obtaining mathematical generality.} A equivalent definition, that
prefigures the Hilbert space notion of a "direct-sum decomposition" of the
space in terms of the eigenspaces of a Hermitian operator ) is a set of
nonempty subsets $\pi=\left\{  B_{1},...,B_{m}\right\}  $ such that every
non-empty subset $S\subseteq U$ can be uniquely represented as the union of a
set of nonempty subsets of the $B_{j}$--in particular $S=\cup\left\{  S\cap
B_{j}\neq\emptyset:j=1,...,m\right\}  $.

As the mathematical tool to describe distinctions versus indistinctions, a
\textit{distinction} or \textit{dit} of $\pi$ is an ordered pair of elements
of $U$ in different blocks of the partition, and the \textit{ditset}
$\operatorname*{dit}\left(  \pi\right)  $ is the set of all the distinctions
of $\pi$ (also called an "apartness relation"). An \textit{indistinction} or
\textit{indit} of $\pi$ is an ordered pair of elements in the same block of
the partition, and the \textit{indit set} $\operatorname*{indit}\left(
\pi\right)  =\cup_{j=1}^{m}B_{j}\times B_{j}$ is the set of all indits of
$\pi$--which is the equivalence relation associated with $\pi$ whose
equivalence classes are the blocks of $\pi$.

The \textit{partial order} on partition is usually defined as $\sigma
\precsim\pi$ (where $\sigma=\left\{  C_{1},...,C_{m^{\prime}}\right\}  $) if
for every $B_{j}\in\pi$, there is a $C_{j^{\prime}}\in\sigma$ such that
$B_{j}\subseteq C_{j^{\prime}}$, but it is easier to just define it by
$\sigma\precsim\pi$ if $\operatorname*{dit}\left(  \sigma\right)
\subseteq\operatorname*{dit}\left(  \pi\right)  $. The join (least upper
bound) and meet (greatest lower bound) operations on partitions on $U$ form
the partition lattice $\Pi\left(  U\right)  $. The most important operation
for our purposes is the join operation where the \textit{join} $\pi\vee\sigma$
is the partition on $U$ whose blocks are the nonempty subsets $B_{j}\cap
C_{j^{\prime}}$ for $j=1,...,m$ and $j^{\prime}=1,...,m^{\prime}$. It could
also be defined using ditsets since: $\operatorname*{dit}\left(  \pi\vee
\sigma\right)  =\operatorname*{dit}\left(  \pi\right)  \cup\operatorname*{dit}%
\left(  \sigma\right)  $. The partition lattice $\Pi\left(  U\right)  $ also
has a top and bottom. The top is the \textit{discrete partition}
$\mathbf{1}_{U}=\left\{  \left\{  u_{1}\right\}  ,...,\left\{  u_{n}\right\}
\right\}  $ with only singleton blocks which makes all possible distinctions,
i.e., $\operatorname*{dit}\left(  \mathbf{1}_{U}\right)  =U\times U-\Delta$
(where $\Delta$ is the diagonal of self-pairs $\left(  u_{i},u_{i}\right)  $).
The bottom is the \textit{indiscrete partition} (or "The Blob"\footnote{Since
$\mathbf{0}_{U}$ is below all other partitions $\pi$ on $U$, it is called "The
Blob" because, as in the Hollywood movie of that name, the Blog absorbs
everything it meets, i.e., $\mathbf{0}_{U}\wedge\pi=\mathbf{0}_{U}$.})
$\mathbf{0}_{U}=\left\{  U\right\}  $ with only one block $U$ and it makes no
distinctions so $\operatorname*{dit}\left(  \mathbf{0}_{U}\right)  =\emptyset$
and $\operatorname*{indit}\left(  \mathbf{0}_{U}\right)  =U\times U$.

There are two notions of `becoming' illustrated as going from the bottom to
top of the Boolean lattice of subsets and the lattice of partitions in terms
of the creation of `its' or dits. The partition notion of becoming is
particularly important for our purposes since it prefigures the notion of
quantum (projective) measurement.%

\begin{center}
\includegraphics[
height=1.5091in,
width=4.0093in
]%
{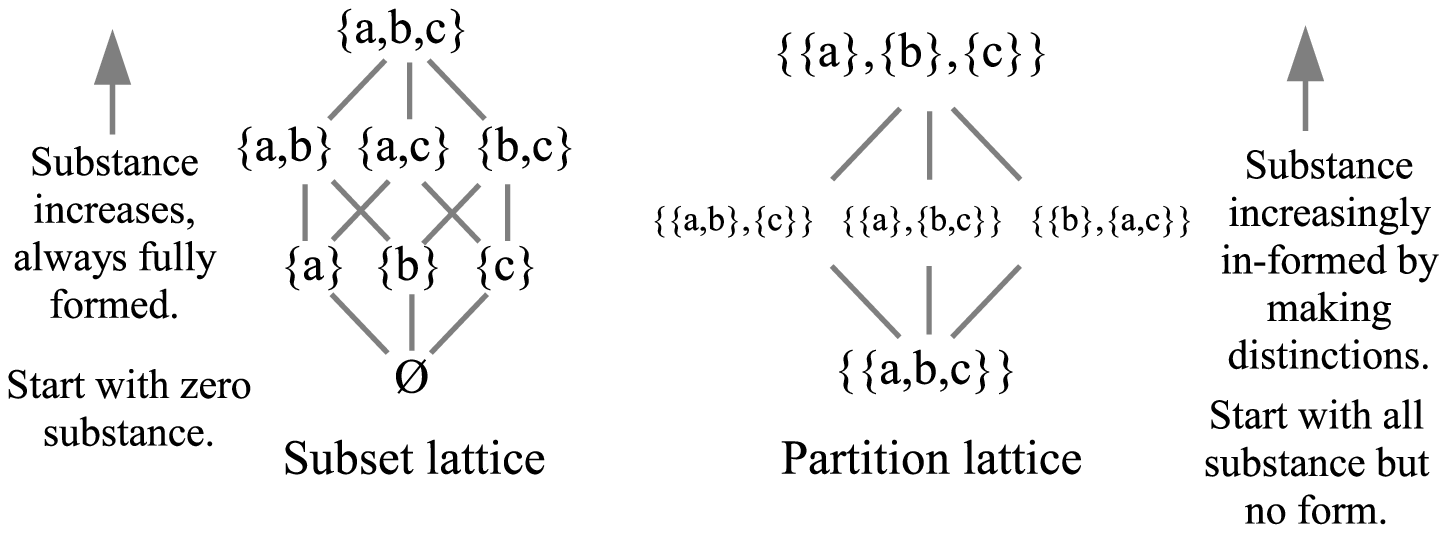}%
\end{center}

\begin{center}
Figure 1: The two dual notions of becoming.
\end{center}

Applied to the `universe,' this gives two stories of creation.

\textit{Subset creation story}: In the Beginning was the Void (no substance)
and then fully definite elements ("Its") were created until the full universe
$U$ was created.

\textit{Partition creation story}: In the Beginning was the Blob--all
substance with no form (i.e., perfect symmetry)--and then, in a Big Bang,
distinctions ("Dits") were created (i.e., symmetries were broken) as the
substance was increasingly in-formed to reach the universe $U$ where
everything is fully distinguished.

\section{The logic of partitions}

The partition join and meet operations were known in the nineteenth century
(e.g., Dedekind and Schr\"{o}der). Ordinary logic is based on the Boolean
logic of subsets of a universe set $U$; propositional logic is the special
case of a one element universe $U$. And subsets (or generally subobjects) are
category-theoretically dual to partitions (or generally quotient objects).
Hence one would naturally expect there to be a dual \textit{logic} of
partitions, but that would require at least the operation of implication on
partitions (corresponding to the Boolean conditional $S\supset T$), but no new
operations on partitions were defined in the twentieth century. As
acknowledged in a 2001 volume commemorating Gian-Carlo Rota: "the only
operations on the family of equivalence relations fully studied, understood
and deployed are the binary join $\vee$ and meet $\wedge$ operations."
\cite[p. 445]{bmp:eqrel} Only in the current century was the implication
operation $\sigma\Rightarrow\pi$ on partitions (which turns the lattice
$\Pi\left(  U\right)  $ into an algebra) defined along with general algorithms
to turn subset logical operations into partition logical operations. The
resulting logic of partitions cemented the notion of a partition as not just a
mathematical concept of combinatorics but a logical concept.
\cite{ell:intropartitions}

There is a parallel development of subset logic and partition logic based on
the dual connection between the elements or "its" of a subset and the
distinctions or "dits" of a partition--which is summarized in Table 1.

\begin{center}%
\begin{tabular}
[c]{l||l|l|}\cline{2-3}%
Its \& Dits & Algebra of subsets$\wp\left(  U\right)  $ of $U$ & Algebra of
partitions $\Pi\left(  U\right)  $ on $U$\\\hline\hline
\multicolumn{1}{|l||}{Its or Dits} & Elements of subsets & Distinctions of
partitions\\\hline
\multicolumn{1}{|l||}{Partial order} & Inclusion of subsets $S\subseteq T$ &
Inclusion of ditsets $\operatorname*{dit}\left(  \sigma\right)  \subseteq
\operatorname*{dit}\left(  \pi\right)  $\\\hline
\multicolumn{1}{|c||}{Logical maps} & \multicolumn{1}{||c|}{Injection
$S\rightarrowtail T$} & \multicolumn{1}{|c|}{Surjection $\pi\twoheadrightarrow
\sigma$}\\\hline
\multicolumn{1}{|l||}{Join} & Union of subsets & Union of ditsets\\\hline
\multicolumn{1}{|l||}{Meet} & Subset of common elements & Ditset of common
dits\\\hline
\multicolumn{1}{|l||}{Top} & Subset $U$ with all elements & Partition
$\mathbf{1}_{U}$ with all distinctions\\\hline
\multicolumn{1}{|l||}{Bottom} & Subset $\emptyset$ with no elements &
Partition $\mathbf{0}_{U}$ with no distinctions\\\hline
Implication & $S\supset T=U$ iff $S\subseteq T$ & $\sigma\Rightarrow
\pi=\mathbf{1}_{U}$ iff $\sigma\precsim\pi$\\\hline
\end{tabular}

Table 1: Elements and Distinctions (Its \& Dits) duality between the two lattices
\end{center}

In subset logic, a formula is valid if for any $U$ ( $\left\vert U\right\vert
\geq1$) and any subsets of $U$ substituted for the atomic variables, the
formula evaluates by the logical subset operations to the top $U$. Similarly
in partition logic, a formula is valid if for any $U$ ($\left\vert
U\right\vert \geq2$) and any partitions on $U$ substituted for the atomic
variables, the formula evaluates by the logical partition operations to the
top $\mathbf{1}_{U}$.

\section{Logical information theory: Logical entropy}

In his writings (and MIT lectures), Gian-Carlo Rota further developed the
parallelism between subsets and partitions by considering their quantitative
versions: "The lattice of partitions plays for information the role that the
Boolean algebra of subsets plays for size or probability." \cite[p.
30]{kung:rota} Since the normalized size of a subset $\Pr\left(  S\right)
=\frac{\left\vert S\right\vert }{\left\vert U\right\vert }$ gives its
Boole-Laplace finite probability, so the "size" of a partition would play a
similar role for information:

\begin{center}
$\frac{\text{Information}}{\text{Partitions}}\approx\frac{\text{Probability}%
}{\text{Subsets}}$.
\end{center}

\noindent Since "Probability is a measure on the Boolean algebra of events"
that gives quantitatively the "intuitive idea of the size of a set", we may
ask by "analogy" for some measure "which will capture some property that will
turn out to be for [partitions] what size is to a set." \cite[p.
67]{rota:fubini} The duality tells us that it is the number of dits in a
partition that gives its size (maximum at the top and minimum at the bottom of
the partition lattice) that is parallel to the number of `its' in a subset
(maximum at the top and minimum at the bottom of the subset lattice).

That is the reasoning that motivates the definition of the \textit{logical
entropy} of a partition as the normalized size of its ditset (equiprobable points):

\begin{center}
$h\left(  \pi\right)  =\frac{\left\vert \operatorname*{dit}\left(  \pi\right)
\right\vert }{\left\vert U\times U\right\vert }=\frac{\left\vert U\times
U\right\vert -\left\vert \operatorname*{indit}\left(  \pi\right)  \right\vert
}{\left\vert U\times U\right\vert }=1-\frac{\cup_{j}\left\vert B_{j}\times
B_{j}\right\vert }{\left\vert U\times U\right\vert }=1-\sum_{j}\left(
\frac{\left\vert B_{j}\right\vert }{\left\vert U\right\vert }\right)  ^{2}$.
\end{center}

\noindent In general, if the points of $U=\left\{  u_{1},...,u_{n}\right\}  $
have general probabilities $p=\left\{  p_{1},...,p_{n}\right\}  $, then
$\Pr\left(  B_{j}\right)  =\sum_{u_{i}\in B_{j}}p_{i}$, so that:

\begin{center}
$h\left(  \pi\right)  =1-\sum_{j=1}^{m}\Pr\left(  B_{j}\right)  ^{2}%
=\sum_{j=1}^{m}\Pr\left(  B_{j}\right)  \left(  1-\Pr\left(  B_{j}\right)
\right)  $.
\end{center}

\noindent The parallelism carries through to the interpretation: $\Pr\left(
S\right)  =\sum_{u_{i}\in S}p_{i}$ is the one-draw probability of getting an
`it' of $S$ and $h\left(  \pi\right)  $ is the two-draw probability of getting
a `dit' of $\pi$.

The logical entropy $h\left(  \pi\right)  $ is a (probability) measure in the
sense of measure theory, i.e.,, $h\left(  \pi\right)  $ is the product measure
$p\times p$ on the ditset $\operatorname*{dit}\left(  \pi\right)  \subseteq
U\times U$. As a measure, the compound notions of joint, conditional, and
mutual logical entropy satisfy the usual Venn diagram relationships. The
well-known Shannon entropy $H\left(  \pi\right)  =\sum_{j=1}^{m}\Pr\left(
B_{j}\right)  \log_{2}\left(  \frac{1}{\Pr\left(  B_{j}\right)  }\right)  $
can also be interpreted in terms of partitions; it is the minimum average
number of binary partitions (bits) it takes to distinguish the blocks of $\pi
$. The Shannon entropy is not a measure in the sense of measure theory but the
compound notions of joint, conditional, and mutual Shannon entropy were
defined so that they satisfy similar Venn-like diagrams. That is possible
because there is a non-linear but monotonic dit-to-bit transform, i.e.,
$1-\Pr\left(  B_{j}\right)  \rightsquigarrow\log_{2}\left(  \frac{1}%
{\Pr\left(  B_{j}\right)  }\right)  $, that takes $h\left(  \pi\right)
=\sum_{j}\Pr\left(  B_{j}\right)  \left(  1-\Pr\left(  B_{j}\right)  \right)
$ to $H\left(  \pi\right)  =\sum_{j}\Pr\left(  B_{j}\right)  \log_{2}\left(
\frac{1}{\Pr\left(  B_{j}\right)  }\right)  $ and which preserves Venn
diagrams. \cite{ell:nf4it}

If a partition $\pi$ is the inverse-image partition $\pi=\left\{
f^{-1}\left(  r\right)  \right\}  _{r\in f\left(  U\right)  }$ of a numerical
attribute $f:U\rightarrow%
\mathbb{R}
$, then $h\left(  \pi\right)  $ is the two-draw probability of getting
different $f$-values. The notion of logical entropy generalizes naturally to
the notion of quantum logical entropy (\cite{ell:nf4it}; \cite{tamir:4open})
where it gives the probability in two independent measurements of the same
state by the same observable that the result gives different eigenvalues.

\section{Partitions as skeletonized quantum states}

There is a very simple way to skeletonize a quantum state to arrive at the
corresponding set notion. Consider $U=\left\{  a,b,c,d\right\}  $ as both a
set of distinct points and also as a orthonormal basis for a $4$-dimensional
Hilbert space. Then a superposition state vector of the form (say)
$\alpha\left\vert a\right\rangle +\beta\left\vert b\right\rangle $ is
\textit{skeletonized} by deleting the complex scalars $\alpha$ and $\beta$,
the Dirac kets, and the addition operation to yield just the set $\left\{
a,b\right\}  $ as a block in a partition. This sets up the skeletal
(many-to-one) dictionary between pure, non-classical mixed states (i.e., mixed
states containing at least one superposition), and the classical mixed
state--with their skeletonized partition versions as in Figure 2--where we
have used the partition shorthand of representing the partition $\left\{
\left\{  a,b\right\}  ,\left\{  c,d\right\}  \right\}  $ as $\left\{
ab,cd\right\}  $ and similarly for the other partitions. The interpretation is
that $a,b,c,d$ are distinct eigenstates of a particle according to some
observable, where "particle" does not mean a classical (or Bohmian) particle
but an entity that can have different levels of objective indefiniteness
(superpositions of the eigenstates such as $a,b,c,$ or $d$) or the
definiteness of an eigenstate.\footnote{The analysis is of standard von
Neumann/Dirac quantum physics, not about quantum field theory where a particle
might be analyzed as a "disturbance in a field."}%

\begin{center}
\includegraphics[
height=2.5452in,
width=5.2736in
]%
{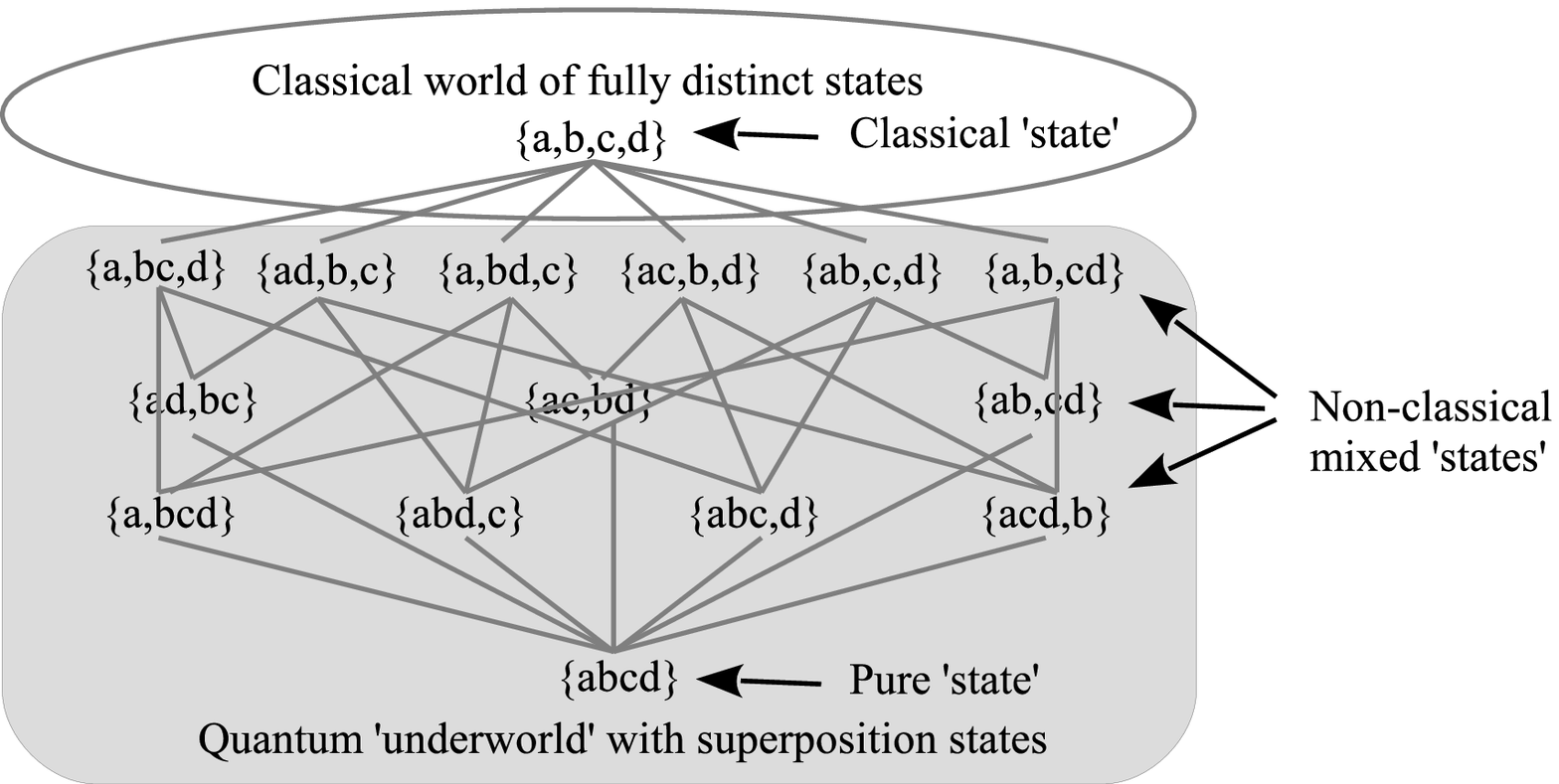}%
\end{center}

\begin{center}
Figure 2: Skeletonized quantum states of a particle as the lattice of partitions.
\end{center}

Figure 2 is the Hasse diagram of the partition lattice on a four element set
which means that each line between two partitions represents the partial order
of refinement with no intermediate partitions. Then the change from a
partition to a more refined one is a "jump." The set level precursor of such a
quantum jump is the "choice function" \cite[p. 60]{halmos:set-theory} which
assigns to each nonempty subset (like a block in a partition) an element of
the subset. That determination of an element is non-deterministic except in
the special case of a singleton set which is the precursor of the quantum
measurement when the outcome has probability one, namely when state being
measured is a single eigenstate of the observable being measured (i.e., a
singleton block in Figure 2).

The top discrete partition $\mathbf{1}_{U}$ is the skeletal version of a
classical mixed state like randomly choosing a leaf in a four-leaf clover or
randomly choosing a `letter' in the four-letter genetic code $U,C,A,G$. One
criterion of classical reality was the idea that it was fully definite or
definite all-the-way-down as in Leibniz's Principle of Identity of
Indistinguishables (PII) \cite[Fourth letter, p. 22]{leib-clarke:letters} or
Kant's Principle of Complete Determination (\textit{omnimoda determinatio}).

\begin{quotation}
\noindent Every thing, however, as to its possibility, further stands under
the principle of thoroughgoing determination; according to which, among all
possible predicates of things, insofar as they are compared with their
opposites, one must apply to it. \cite[B600]{kant:cpr}
\end{quotation}

\noindent Thus two distinct things must have some predicate to distinguish
them or if there is no way to distinguish them, then they are the same thing.
This principle of classicality characterizes the `classical' state
$\mathbf{1}_{U}$:

\begin{center}
For any $u,u^{\prime}\in U$, if $\left(  u,u^{\prime}\right)  \in
\operatorname*{indit}\left(  \mathbf{1}_{U}\right)  $, then $u=u^{\prime}$

Partition logical version of Principle of Identity of Indistinguishables.
\end{center}

\noindent Any non-classical state in the skeletal representation is a
partition $\pi$ with a non-singleton `superposition' block, e.g., $\left\{
a,b\right\}  $ so $\left(  a,b\right)  \in\operatorname*{indit}\left(
\pi\right)  $ but $a\neq b$. As noted above, any idealized measurement of a
classical state (i.e., a singleton) gives the outcome of that state with
probability one.

In addition to PII, Leibniz had other metaphysical principles characteristic
of the classical notion of reality. His Principle of Continuity was expressed
by "\textit{Natura non facit saltus}" (Nature does not make jumps) \cite[Bk.
IV, chap. xvi]{leibniz:ne} and his Principle of Sufficient Reason was
expressed as the statement "that nothing happens without a reason why it
should be so rather than otherwise" \cite[Second letter, p. 7]%
{leib-clarke:letters}. All these classical principles are violated in the
quantum world; PII is violated by superpositions of bosons, Continuity is
violated by the quantum jumps, and Sufficient Reason is violated by the
objective probabilities of QM.

This skeletal representation of quantum states is summarized in Table 2.

\begin{center}%
\begin{tabular}
[c]{|c|c|}\hline
Partition concept & Corresponding quantum concept\\\hline\hline
Non-singleton block, e.g., $\left\{  a,b\right\}  $ & Superposition pure
state\\\hline
Indiscrete partition $\mathbf{0}_{U}=\left\{  \left\{  a,b,c,d\right\}
\right\}  $ & Largest pure state\\\hline
Singleton block, e.g., $\left\{  d\right\}  $ & Classical state (no
superposition)\\\hline
Discrete partition $\mathbf{1}_{U}=\left\{  \left\{  a\right\}  ,\left\{
b\right\}  ,\left\{  c\right\}  ,\left\{  d\right\}  \right\}  $ & Classical
mixture of states\\\hline
Partition, e.g., $\left\{  \left\{  a,b,c\right\}  ,\left\{  d\right\}
\right\}  $ & Mixture of orthogonal states\\\hline
\end{tabular}

Table 2: Corresponding partition and quantum concepts
\end{center}

\section{Superposition as indefiniteness in the quantum `underworld'}

The biggest `enemy' to understanding QM is the wave imagery, not to mention
the name "wave mechanics." That imagery interprets superposition as the sum of
two definite waves to give another \textit{definite} wave as in Figure 3.%

\begin{center}
\includegraphics[
height=1.7997in,
width=2.7129in
]%
{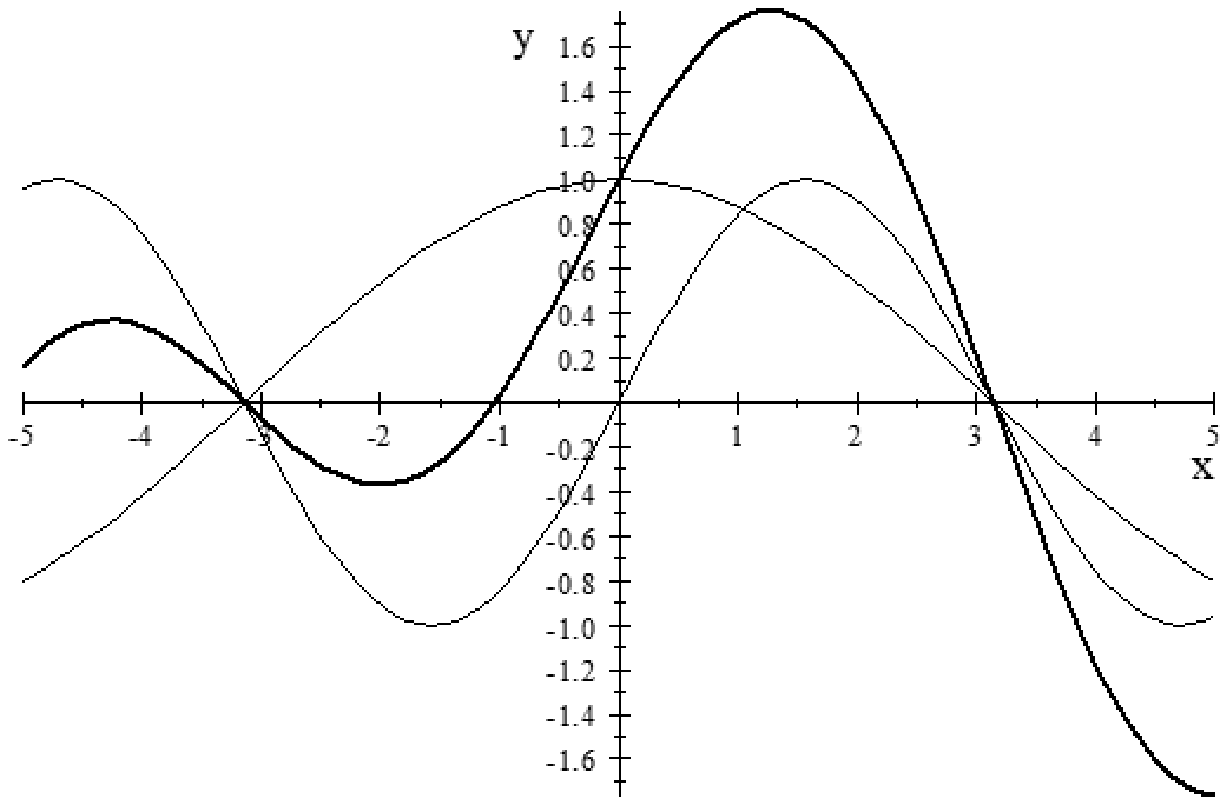}%
\end{center}

\begin{center}
Figure 3: `Wrong' image of superposition in QM as sum of waves
\end{center}

It is the wave \textit{imagery} that is wrong, not the math of waves since any
vector in a space over the complex numbers $%
\mathbb{C}
$ automatically has a wave imagery in the polar representation as having an
amplitude and phase. The wave interpretation is a misleading artefact of the
use of complex numbers in the math of QM, which is because (among other
reasons) they are algebraically complete so that the observable operators will
have a complete set of eigenvectors \cite[p. 67, fn. 7]{weinberg: lectures},
not because the `wave function' describes any physical waves.\footnote{In view
of the century-long difficulties in interpreting QM as "wave mechanics,"
Einstein's statement: "The Lord is subtle, but not malicious," may be too
optimistic.} It is a fact of the mathematics that the addition of vectors in a
vector space over $%
\mathbb{C}
$ can \textit{always} be represented as the superposition of waves with
interference effects.

\begin{quotation}
\noindent The wave formalism offers a convenient mathematical representation
of this latency, for not only can the mathematics of wave effects, like
interference and diffraction, be expressed in terms of the addition of vectors
(that is, their linear superposition; see \cite[Chap. 29.5]{feynman:vI}), but
the converse, also holds. \cite[p. 303]{hughes:structure}
\end{quotation}

On the partitional (insstead of wave) approach, a superposition of two
definite states $\left\{  a\right\}  $ and $\left\{  b\right\}  $ is a state
$\left\{  a,b\right\}  $ that is indefinite between the two definite states.
For a pictorial image, Figure 4 gives the superposition of two isosceles
triangles with labelled edges and vertices as the triangle that is indefinite
on the edges and vertices where they differ and only definite where the two
triangles are the same--and certainly not the triangle that is doubly definite
(like a double-exposure photograph).%

\begin{center}
\includegraphics[
height=1.0438in,
width=3.6426in
]%
{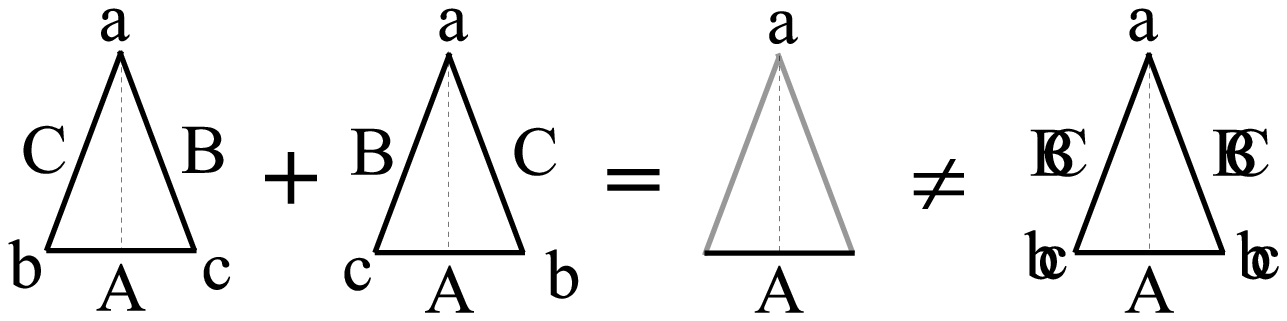}%
\end{center}

\begin{center}
Figure 4: Imagery of superposition as indefiniteness
\end{center}

\noindent The fact that where superposed states have the same property is
still definite (like vertex label $a$, the side label $A$) in the
superposition is a little noticed fact about quantum superpositions.

\begin{quotation}
\noindent It follows from the linearity of the operators which represent
observables of quantum mechanical systems that any measurable physical
property which happens to be shared by all of the individual mathematical
terms of some particular superposition (written down in any particular basis)
will necessarily also be shared by the full superposition, considered as a
single quantum-mechanical state, as well. \cite[p. 234]{albert:superposition}
\end{quotation}

\noindent This means that the notion of superposition in QM and the notion of
abstraction (e.g., in mathematics) are `essentially' the same notion viewed
from different angles \cite{ell:abstr}. In a superposition, the emphasis is on
the indefiniteness resulting from where the elements of a set (i.e., a
non-singleton equivalence class in an equivalence relation) differ like the
vertex labels $b$ and $c$, and the side labels $B$ and $C$ in Figure 4, while
in abstraction, the emphasis is on the properties that are the same for the
elements in the set like the vertex label $a$ and side label $A$. It is like
the different viewpoints of seeing a glass as half-empty or as half-full.

It is not a new idea that there is a quantum `underworld' of indefinite
superposition states `beneath' the classical space-time world of definite
states. That view was previously expressed in the language of the quantum
world of potentialities (or latencies) versus the classical world of
actualities (\cite{heisen:phy-phil}; \cite{margenau:1954}; \cite[Sec.
10.2]{hughes:structure}; \cite{shim:vienna}; \cite{fleming:actual-pot};
\cite{kastner-et-al:potentia}, \cite{kastner:tiandposs};
\cite{deronde:superpos}).

\begin{quotation}
\noindent Heisenberg \cite[p. 53]{heisen:phy-phil}... used the term
"potentiality" to characterize a property which is objectively indefinite,
whose value when actualized is a matter of objective chance, and which is
assigned a definite probability by an algorithm presupposing a definite
mathematical structure of states and properties. Potentiality is a modality
that is somehow intermediate between actuality and mere logical possibility.
That properties can have this modality, and that states of physical systems
are characterized partially by the potentialities they determine and not just
by the catalogue of properties to which they assign definite values, are
profound discoveries about the world, rather than about human knowledge.
\cite[p. 6]{shim:vienna}
\end{quotation}

Ruth Kastner even uses the imagery of an iceberg \cite[p. 3]%
{kastner:tiandposs} with the classical world above water and the quantum world
beneath the water--an imagery that is filled out by the Figure 2 imagery of
the partition lattice as the skeletonized classical and quantum states. The
language of "potentialities" or the Aristotelean notion of "potentia" is not
very felicitous since they are taken to be realities, not mere possibilities.
Hence some prefer the language of "latencies" (e.g., Henry Margenau and R. I.
G. Hughes), but in both the cases of "potentialities" and "latencies," the key
idea is objective indefiniteness.

\begin{quotation}
\noindent The historical reference should perhaps be dismissed, since quantum
mechanical potentiality is completely devoid of teleological significance,
which is central to Aristotle's conception. What it has in common with
Aristotle's conception is the indefinite character of certain properties of
the system. \cite[pp. 313-4]{shim:vol2}
\end{quotation}

\noindent And Margenau notes that the measurement of observables "forces them
out of indiscriminacy or latency" \cite[p. 10]{margenau:1954}--which indicates
that Margenau also interprets "latency" in terms of indeterminacy or
indefiniteness. Kastner also considers the indeterminacy of values as a key
characteristic of the real potentia \cite[p. 3]{kastner:tiandposs}.

\section{Quantum states}

To demonstrate our thesis that the math of QM is the Hilbert space version of
the math of partitions, we need to first focus on the three main concepts in
the math of QM: 1) the quantum state, 2) the quantum observable, and 3) the
quantum ( always projective) measurement.

A quantum state can be presented either as a state vector or as a density
matrix \cite{weinberg:densitym}. The density matrix approach best displays the
relevant information for the partitional interpretation. Hence we start by
transferring the structure of a partition $\pi$ into its density matrix form
$\rho\left(  \pi\right)  $. The initial data is a partition $\pi=\left\{
B_{1},...,B_{m}\right\}  $ on $U$ with (positive) point probabilities
$p=\left(  p_{1},...,p_{n}\right)  $. For each $B_{j}\in\pi$, define
$\left\vert b_{j}\right\rangle $ as the column vector with the $i^{th}$ entry
being $\sqrt{p_{i}/\Pr\left(  B_{j}\right)  }$ if $u_{i}\in B_{j}$, else $0$
so that $\left\langle b_{j^{\prime}}|b_{j}\right\rangle =\delta_{jj^{\prime}}%
$. We form the projection matrix $\rho\left(  B_{j}\right)  =\left\vert
b_{j}\right\rangle \left\langle b_{j}\right\vert $ with the $i,k$-entry being
$\rho\left(  B_{j}\right)  _{ik}=\frac{\sqrt{p_{i}p_{k}}}{\Pr\left(
B_{j}\right)  }$ if $u_{i},u_{k}\in B_{j}$, else $0$. Then the density matrix
$\rho\left(  \pi\right)  $ is the probability sum of these projectors:

\begin{center}
$\rho\left(  \pi\right)  =\sum_{j=1}^{m}\Pr\left(  B_{j}\right)  \left\vert
b_{j}\right\rangle \left\langle b_{j}\right\vert $.
\end{center}

\noindent Then it is easily checked that $\rho\left(  \pi\right)  _{ik}%
=\sqrt{p_{i}p_{k}}$ if $\left(  u_{i},u_{k}\right)  \in\operatorname*{indit}%
\left(  \pi\right)  $, else $0$. Thus the non-zero entries in $\rho\left(
\pi\right)  $ represent the indits of $\pi$ and the zero entries represent the
distinctions of $\pi$. A density matrix is not only Hermitian but positive so
its eigenvalues are non-negative real numbers $\lambda_{i}$ which sum to $1$,
i.e., $\sum_{i=1}^{n}\lambda_{i}=1$. In the case of $\rho\left(  \pi\right)
$, there are $m$ non-zero eigenvalues $\Pr\left(  B_{j}\right)  $ with the
remaining $n-m$ eigenvalues of $0$.

For the classical state $\mathbf{1}_{U}$, its density matrix $\rho\left(
\mathbf{1}_{U}\right)  $ is a diagonal matrix with the point probabilities
along the diagonal, e.g. "the statistical mixture describing the state of a
classical dice before the outcome of the throw\textquotedblright\ \cite[p.
176]{auletta:qm}. Thus the non-classical states in the skeletal representation
are the ones where $\rho\left(  \pi\right)  $ has non-zero off-diagonal
elements indicating the `amplitudes' $\sqrt{p_{i}p_{k}}$ of the corresponding
diagonal states ($u_{i}$ and $u_{k}$) blobbing or cohering together in a
superposition. Since superposition states are the key non-classical states, it
is these non-zero off-diagonal "coherences" \cite[p. 303]%
{cohen-tennoudji:vol1-2} that account for the non-classical interference
effects in the Hilbert space version.

\begin{quotation}
\noindent\lbrack T]he off-diagonal terms of a density matrix ... are often
called quantum coherences because they are responsible for the interference
effects typical of quantum mechanics that are absent in classical dynamics.
\cite[p. 177]{auletta:qm}.
\end{quotation}

In the full non-skeletal Hilbert space case of a density matrix $\rho$, it has
a spectral decomposition $\rho=\sum_{i=1}^{n}\lambda_{i}\left\vert
u_{i}\right\rangle \left\langle u_{i}\right\vert $ with an orthonormal basis
$\left\{  \left\vert u_{i}\right\rangle \right\}  _{i=1}^{n}$so $\left\langle
u_{i^{\prime}}|u_{i}\right\rangle =\delta_{ii^{\prime}}$ and where the
non-negative eigenvalues $\lambda_{i}$ sum to $1$. Thus for the concept of a
quantum state, we have the skeletal set level presentation of a partition and
the corresponding Hilbert space version of that partition math as summarized
in Table 3.

\begin{center}%
\begin{tabular}
[c]{l||l|l|}\cline{2-3}%
Quantum States & Partition math & Hilbert space math\\\hline\hline
\multicolumn{1}{|l||}{Density matrix} & $\rho\left(  \pi\right)  $ & $\rho
$\\\hline
\multicolumn{1}{|l||}{ON vectors} & $\left\langle b_{j^{\prime}}%
|b_{j}\right\rangle =\delta_{jj^{\prime}}$ & $\left\langle u_{i^{\prime}%
}|u_{i}\right\rangle =\delta_{ii^{\prime}}$\\\hline
\multicolumn{1}{|l||}{Non-negative eigenvalues} & \multicolumn{1}{||c|}{$\Pr
\left(  B_{1}\right)  ,...,\Pr\left(  B_{m}\right)  ,0,...,0$} & $\lambda
_{1},...,\lambda_{n}$\\\hline
\multicolumn{1}{|l||}{Spectral decomposition} & $\rho\left(  \pi\right)
=\sum_{j=1}^{m}\Pr\left(  B_{j}\right)  \left\vert b_{j}\right\rangle
\left\langle b_{j}\right\vert $ & $\rho=\sum_{i=1}^{n}\lambda_{i}\left\vert
u_{i}\right\rangle \left\langle u_{i}\right\vert $\\\hline
\multicolumn{1}{|l||}{Non-zero off-diagonal entries} & Cohering of diag.
elements & Coherence of diag. elements\\\hline
\end{tabular}

Table 3: Quantum state $\rho$ as Hilbert space version of partition
$\rho\left(  \pi\right)  $
\end{center}

\section{Quantum observables}

We have seen how the notion of a quantum state was prefigured at the set level
by (i.e., has the set level precursor of) a partition on a set with point
probabilities. In a similar manner, the notion of a quantum observable is
prefigured at the set level by the inverse-image partition $\left\{
f^{-1}\left(  r\right)  \right\}  _{r\in f\left(  U\right)  }$ of a real-value
numerical attribute $f:U\rightarrow%
\mathbb{R}
$.

In the folklore of mathematics, there is a semi-algorithmic procedure to
connect set concepts with the corresponding vector (Hilbert) space concepts.
We will call it the "Yoga of Linearization":

\begin{center}
For any given set-concept, apply it to a basis set of a vector space

and whatever is linearly generated is the corresponding vector space concept.

\textit{The Yoga of Linearization}.
\end{center}

\noindent To apply the Yoga, we first take $U$ as just a universal set and
consider some set-based concept, and then we consider $U$ as a basis set
(e.g., ON basis of a Hilbert space) of a vector space $V$ and see what is
linearly generated. For instance, a subset $S$ of a basis set $U$ generates a
subspace $\left[  S\right]  $ of the space $V$. The cardinality of the subset
gives the dimension of the subspace. A real-valued numerical attribute
$f:U\rightarrow%
\mathbb{R}
$ defines a Hermitian operator $F:V\rightarrow V$ (where $V=\left[  U\right]
$) by defining $F$ on the basis set $U$ as $Fu=f\left(  u\right)  u$ or, using
the fancier notation, $F\left\vert u\right\rangle =f\left(  u\right)
\left\vert u\right\rangle $. To better analyze the numerical attribute, let
$f\upharpoonright S=rS$ stand for "the value of $f$ on the subset $S$ is $r$".
That is the set level version of the eigenvalue/eigenvector equation
$F\left\vert u\right\rangle =r\left\vert u\right\rangle $. Hence we see that
the set version of an eigenvector is a constant set $S$ of $f$ and the set
version of an eigenvalue of an eigenvector is the constant value $r$ on a
constant set $S$. A characteristic function $\chi_{S}:U\rightarrow\left\{
0,1\right\}  \subseteq%
\mathbb{R}
$ has only two constant sets $S=\chi_{S}^{-1}\left(  1\right)  $ and
$S^{c}=U-S=\chi_{S}^{-1}\left(  0\right)  $. The Yoga yields the corresponding
vector space notion which is a projection operator $P_{\left[  S\right]  }$
which is defined by $P_{\left[  S\right]  }\left\vert u\right\rangle
=\left\vert u\right\rangle $ if $u\in S$, else $\mathbf{0}$ (zero vector)
which has the eigenvalues of $0$ and $1$. A Hermitian (or self-adjoint)
operator $F$ in QM has spectral decomposition $F=\sum_{\lambda_{i}}\lambda
_{i}P_{V_{i}}$ where the sum is over the real eigenvalues $\lambda_{i}$ and
the projections $P_{V_{i}}$ to their eigenspaces. Bearing in mind the
correlation given by the Yoga, we can define the `spectral decomposition' of
the numerical attribute $f:U\rightarrow%
\mathbb{R}
$ as $f=\sum_{r\in f\left(  U\right)  }r\chi_{f^{-1}\left(  r\right)  }$.
Starting at the quantum level with a Hermitian operator $F:V\rightarrow V$ and
a basis set $U$ of eigenvectors of $F$, then $f$ is obtained as the eigenvalue
function  $f:U\rightarrow%
\mathbb{R}
$.

An important application of the Yoga is to the notion of a set partition
$\pi=\left\{  f^{-1}\left(  r\right)  \right\}  _{r\in f\left(  U\right)  }$
as the inverse-image of a numerical attribute. Applied to the basis set
$\left\{  \left\vert u_{i}\right\rangle \right\}  _{i=1}^{n}$ used to define
$F$ by $F\left\vert u_{i}\right\rangle =f\left(  u_{i}\right)  \left\vert
u_{i}\right\rangle $, each block $f^{-1}\left(  r\right)  $ of $\pi$ generates
the eigenspace of eigenvectors for the eigenvalue $r$. This eigenspaces
$\left\{  V_{r}\right\}  _{r\in f\left(  U\right)  }$ form a
\textit{direct-sum decomposition} (DSD) of $V$, i.e., $V=\oplus_{r\in f\left(
U\right)  }V_{r}$, where a DSD is defined as a set of non-zero subspaces
$\left\{  V_{r}\right\}  _{r\in f\left(  U\right)  }$ such that every non-zero
vector $v\in V$ has a unique representation as a sum $v=\sum_{r\in f\left(
U\right)  }v_{r}$ of vectors $v_{r}\in V_{r}$. It was noted previously that a
set partition $\pi=\left\{  B_{1},...,B_{m}\right\}  $ has a similar
definition since every non-empty subset $S$ has a unique representation as the
union of subsets of the $\left\{  B_{j}\right\}  _{j=1}^{m}$. If the union of
the $B_{j}$'s was not all of $U$, then $U-\cup_{j=1}^{m}B_{j}$ would have no
representation, and if $S=B_{j}\cap B_{j^{\prime}}\neq\emptyset$, then $S$ has
two representations as subsets of the $B_{j}$'s. Moreover, $S\cap():\wp\left(
U\right)  \rightarrow\wp\left(  U\right)  $ is a projection operator that
takes any subset $T\in\wp\left(  U\right)  $ to $S\cap T\in\wp\left(
U\right)  $. Thus we have $\cup_{r\in f\left(  U\right)  }\left(
f^{-1}\left(  r\right)  \cap()\right)  =I:\wp\left(  U\right)  \rightarrow
\wp\left(  U\right)  $ whose quantum version is resolution of unity
$\sum_{r\in f\left(  U\right)  }P_{V_{r}}=I:V\rightarrow V$. And lastly, we
might apply the Yoga to the Cartesian product $U\times U^{\prime}$ where $U$
and $U^{\prime}$ are basis sets for $V$ and $V^{\prime}$. Then the ordered
pairs $\left(  u,u^{\prime}\right)  \in U\times U^{\prime}$ (bi)linearly
generate the tensor product $V\otimes V^{\prime}$ where the ordered pair
$\left(  u,u^{\prime}\right)  $ is customarily written as $u\otimes u^{\prime
}$or $\left\vert u\right\rangle \otimes\left\vert u^{\prime}\right\rangle $.

These results of the Yoga of Linearization are summarized in Table 4.

\begin{center}%
\begin{tabular}
[c]{|c|c|}\hline
Set concept (skeletons) & Vector-space concept\\\hline\hline
{\small Partition} $\left\{  f^{-1}\left(  r\right)  \right\}  _{r\in f\left(
U\right)  }$ & {\small DSD }$\left\{  V_{r}\right\}  _{r\in f\left(  U\right)
}$\\\hline
$U=\uplus_{r\in f\left(  U\right)  }f^{-1}\left(  r\right)  $ & $V=\oplus
_{r\in f\left(  U\right)  }V_{r}$\\\hline
Numerical attribute $f:U\rightarrow%
\mathbb{R}
$ & Observable $Fu_{i}=f\left(  u_{i}\right)  u_{i}$\\\hline
$f\upharpoonright S=rS$ & $Fu_{i}=ru_{i}$\\\hline
Constant set $S$\ of $f$ & Eigenvector $u_{i}$\ of $F$\\\hline
Value $r$\ on constant set $S$ & Eigenvalue $r$\ of eigenvector $u_{i}%
$\\\hline
Characteristic fcn. $\chi_{S}:U\rightarrow\left\{  0,1\right\}  $ & Projection
operator $P_{\left[  S\right]  }u_{i}=\chi_{S}(u_{i})u_{i}$\\\hline
$\cup_{r\in f\left(  U\right)  }\left(  f^{-1}\left(  r\right)  \cap()\right)
=I:\wp\left(  U\right)  \rightarrow\wp\left(  U\right)  $ & $\sum_{r\in
f\left(  U\right)  }P_{V_{r}}=I:V\rightarrow V$\\\hline
Spectral Decomp. $f=\sum_{r\in f\left(  U\right)  }r\chi_{f^{-1}\left(
r\right)  }$ & Spectral Decomp. $F=\sum_{r\in f\left(  U\right)  }rP_{V_{r}}%
$\\\hline
Set of $r$-constant sets $\wp\left(  f^{-1}\left(  r\right)  \right)  $ &
Eigenspace $V_{r}$ of $r$-eigenvectors\\\hline
Cartesian product $U\times U^{\prime}$ & Tensor product $V\otimes V^{\prime}%
$\\\hline
\end{tabular}

Table 4: Skeletal set-level concepts and the corresponding vector (Hilbert)
space concepts
\end{center}

\section{Quantum Measurement}

The third basic concept to be analyzed is quantum measurement (always
projective). The connection between the set level notion of measurement and
the quantum level is the L\"{u}ders mixture operation (\cite{luders:meas};
\cite[p. 279]{auletta:qm}) that can be applied at both levels. At the set
level, we have the skeletal state represented by a density matrix $\rho\left(
\pi\right)  $ and we have an `observable' or real-value numerical attribute,
say, $g:U\rightarrow%
\mathbb{R}
$ whose inverse-image is the partition $\sigma=\left\{  g^{-1}\left(
r\right)  \right\}  _{r\in g\left(  U\right)  }$. The L\"{u}ders mixture
operation applies the `observable' to the density matrix $\rho\left(
\pi\right)  $ to arrive at the post-measurement density matrix $\hat{\rho
}\left(  \pi\right)  $. The operation uses the $n\times n$ projection matrices
for the blocks $g^{-1}\left(  r\right)  $ of $\sigma$ which are diagonal
matrices $P_{g^{-1}\left(  r\right)  }$ whose diagonal elements are the values
of the characteristic function $\chi_{g^{-1}\left(  r\right)  }$. Then the
post-measurement density matrix is:

\begin{center}
$\hat{\rho}\left(  \pi\right)  =\sum_{r\in g\left(  U\right)  }P_{g^{-1}%
\left(  r\right)  }\rho\left(  \pi\right)  P_{g^{-1}\left(  r\right)  }$

Set version of L\"{u}ders mixture operation.
\end{center}

Then it is an easy result:

\noindent\textbf{Theorem}: $\hat{\rho}\left(  \pi\right)  =\rho\left(  \pi
\vee\sigma\right)  $.

\noindent Thus the set version of quantum level projective measurement in the
math of QM is the partition join operation where $\operatorname*{dit}\left(
\pi\vee\sigma\right)  =\operatorname*{dit}\left(  \pi\right)  \cup
\operatorname*{dit}\left(  \sigma\right)  $ and, by DeMorgan's Law,
$\operatorname*{indit}\left(  \pi\vee\sigma\right)  =\operatorname*{indit}%
\left(  \pi\right)  \cap\operatorname*{indit}\left(  \sigma\right)  $.

\textbf{Example}: Let $\pi=\left\{  \left\{  a\right\}  ,\left\{  b,c\right\}
\right\}  $ with probabilities $\Pr\left(  \left\{  a\right\}  \right)
=\frac{1}{3}$, $\Pr\left(  \left\{  b\right\}  \right)  =\frac{1}{4}$, and
$\Pr\left(  \left\{  c\right\}  \right)  =\frac{5}{12}$ in $U=\left\{
a,b,c\right\}  $ and $\sigma=\left\{  \left\{  a,b\right\}  ,\left\{
c\right\}  \right\}  $ so $\sigma=g^{-1}$ for any $g:U\rightarrow%
\mathbb{R}
$ that assigns the same $g$-value to $a$ and $b$ with a different value for
$c$. Then the density matrix for $\pi$ and the projections matrices for the
blocks of $\sigma$ are:

\begin{center}
$\rho\left(  \pi\right)  =%
\begin{bmatrix}
\frac{1}{3} & 0 & 0\\
0 & \frac{1}{4} & \frac{\sqrt{5}}{4\sqrt{3}}\\
0 & \frac{\sqrt{5}}{4\sqrt{3}} & \frac{5}{12}%
\end{bmatrix}
$, $P_{\left\{  a,b\right\}  }=%
\begin{bmatrix}
1 & 0 & 0\\
0 & 1 & 0\\
0 & 0 & 0
\end{bmatrix}
$, and $P_{\left\{  c\right\}  }=%
\begin{bmatrix}
0 & 0 & 0\\
0 & 0 & 0\\
0 & 0 & 1
\end{bmatrix}
$.
\end{center}

\noindent The L\"{u}ders mixture operation is:

\begin{center}
$\hat{\rho}\left(  \pi\right)  =P_{\left\{  a,b\right\}  }\rho\left(
\pi\right)  P_{\left\{  a,b\right\}  }+P_{\left\{  c\right\}  }\rho\left(
\pi\right)  P_{\left\{  c\right\}  }$

$=%
\begin{bmatrix}
1 & 0 & 0\\
0 & 1 & 0\\
0 & 0 & 0
\end{bmatrix}%
\begin{bmatrix}
\frac{1}{3} & 0 & 0\\
0 & \frac{1}{4} & \frac{\sqrt{5}}{4\sqrt{3}}\\
0 & \frac{\sqrt{5}}{4\sqrt{3}} & \frac{5}{12}%
\end{bmatrix}%
\begin{bmatrix}
1 & 0 & 0\\
0 & 1 & 0\\
0 & 0 & 0
\end{bmatrix}
$

$+%
\begin{bmatrix}
0 & 0 & 0\\
0 & 0 & 0\\
0 & 0 & 1
\end{bmatrix}%
\begin{bmatrix}
\frac{1}{3} & 0 & 0\\
0 & \frac{1}{4} & \frac{\sqrt{5}}{4\sqrt{3}}\\
0 & \frac{\sqrt{5}}{4\sqrt{3}} & \frac{5}{12}%
\end{bmatrix}%
\begin{bmatrix}
0 & 0 & 0\\
0 & 0 & 0\\
0 & 0 & 1
\end{bmatrix}
$

$=\allowbreak\allowbreak%
\begin{bmatrix}
\frac{1}{3} & 0 & 0\\
0 & \frac{1}{4} & 0\\
0 & 0 & 0
\end{bmatrix}
+%
\begin{bmatrix}
0 & 0 & 0\\
0 & 0 & 0\\
0 & 0 & \frac{5}{12}%
\end{bmatrix}
=\allowbreak%
\begin{bmatrix}
\frac{1}{3} & 0 & 0\\
0 & \frac{1}{4} & 0\\
0 & 0 & \frac{5}{12}%
\end{bmatrix}
$.
\end{center}

\noindent Since $\pi\vee\sigma=\left\{  \left\{  a\right\}  ,\left\{
b\right\}  ,\left\{  c\right\}  \right\}  =\mathbf{1}_{U}$, we see that the
post-measurement density matrix is $\hat{\rho}\left(  \pi\right)  =\rho\left(
\pi\vee\sigma\right)  =\rho\left(  \mathbf{1}_{U}\right)  $. Thus the
superposition of $\left\{  b\right\}  $ and $\left\{  c\right\}  $ in $\pi$
got distinguished since $b$ and $c$ had different $g$-values. The new
distinctions in $\operatorname*{dit}\left(  \sigma\right)
-\operatorname*{dit}\left(  \pi\right)  $ are $\left(  b,c\right)  $ (along
with $\left(  c,b\right)  $) and those were the non-zero off-diagonal elements
(coherences) of $\rho\left(  \pi\right)  $ that got zeroed (distinguished or
decohered) in $\rho\left(  \pi\vee\sigma\right)  $. Figure 5 shows the join of
$\left\{  \left\{  a\right\}  ,\left\{  b,c\right\}  \right\}  $ and $\left\{
\left\{  a,b\right\}  ,\left\{  c\right\}  \right\}  $ is their least upper
bound $\mathbf{1}_{U}=\left\{  \left\{  a\right\}  ,\left\{  b\right\}
,\left\{  c\right\}  \right\}  $.%

\begin{center}
\includegraphics[
height=1.471in,
width=2.258in
]%
{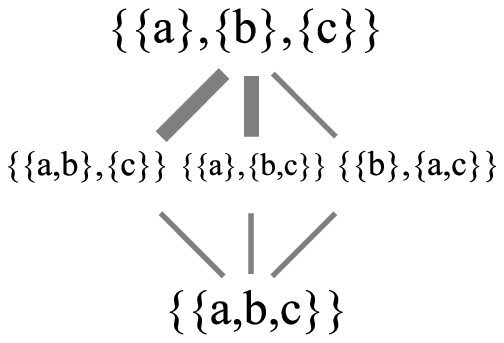}%
\end{center}

\begin{center}
Figure 5: Partition lattice with join $\left\{  \left\{  a\right\}  ,\left\{
b,c\right\}  \right\}  \vee\left\{  \left\{  a,b\right\}  ,\left\{  c\right\}
\right\}  =\left\{  \left\{  a\right\}  ,\left\{  b\right\}  ,\left\{
c\right\}  \right\}  $
\end{center}

The L\"{u}ders mixture is not the end of the measurement process. The
measurement returns one of the $g$-values, say the (degenerate) one for
$\left\{  a,b\right\}  $. Then the L\"{u}ders Rule \cite[Appendix
B]{hughes:structure} gives the final density matrix which is the corresponding
term in the L\"{u}ders mixture sum, e.g., $P_{\left\{  a,b\right\}  }%
\rho\left(  \pi\right)  P_{\left\{  a,b\right\}  }$, normalized so the final
density matrix is the (in this case, classical) mixed state:

\begin{center}
$\frac{P_{\left\{  a,b\right\}  }\rho\left(  \pi\right)  P_{\left\{
a,b\right\}  }}{\operatorname*{tr}\left[  P_{\left\{  a,b\right\}  }%
\rho\left(  \pi\right)  P_{\left\{  a,b\right\}  }\right]  }=%
\begin{bmatrix}
\frac{1}{3} & 0 & 0\\
0 & \frac{1}{4} & 0\\
0 & 0 & 0
\end{bmatrix}
\frac{1}{7/12}=%
\begin{bmatrix}
\frac{4}{7} & 0 & 0\\
0 & \frac{3}{7} & 0\\
0 & 0 & 0
\end{bmatrix}
$.
\end{center}

In the general Hilbert space case, the Hermitian operator $G$ is given by its
DSD of eigenspaces $\left\{  V_{r}\right\}  _{r\in g\left(  U\right)  }$
(where $g:U\rightarrow%
\mathbb{R}
$ is the eigenvalue function assigning the appropriate real eigenvalue to each
vector in an ON basis $U$ of eigenvectors of $G$). The state being measured is
given by density matrix $\rho$ expressed in the ON basis $U$, and the
L\"{u}ders mixture operation uses the projection matrices $P_{V_{r}}$ to the
eigenspaces of $G$ to determine the post-measurement density matrix $\hat
{\rho}$. The Hilbert space version of the set operation is:

\begin{center}
$\hat{\rho}=\sum_{r\in g\left(  U\right)  }P_{V_{r}}\rho P_{V_{r}}$

Hilbert space version of L\"{u}ders mixture operation.
\end{center}

We saw previously how the notion of logical entropy (and its quantum
counterpart) was based on the notion of a quantitative measure of distinctions
of a partition. Hence logical entropy is the \textit{natural} notion to
measure the changes in a density matrix under a measurement. For instance, in
the above example where $\pi=\left\{  \left\{  a\right\}  ,\left\{
b,c\right\}  \right\}  $, the block probabilities are $\Pr\left(  \left\{
a\right\}  \right)  =\frac{1}{3}$ and $\Pr\left(  \left\{  b,c\right\}
\right)  =\frac{2}{3}$, and the logical entropy is: $h\left(  \pi\right)
=1-\Pr\left(  \left\{  a\right\}  \right)  ^{2}-\Pr\left(  \left\{
b,c\right\}  \right)  ^{2}=1-\frac{1}{9}-\frac{4}{4}=\frac{4}{9}$. When the
partition $\pi$ is represented as a density matrix $\rho\left(  \pi\right)  $,
then the logical entropy could also be computed as:

\begin{center}
$h\left(  \pi\right)  =h\left(  \rho\left(  \pi\right)  \right)
=1-\operatorname*{tr}\left[  \rho\left(  \pi\right)  ^{2}\right]
=1-\operatorname*{tr}\allowbreak%
\begin{bmatrix}
\frac{1}{9} & 0 & 0\\
0 & \frac{1}{6} & \frac{1}{18}\sqrt{3}\sqrt{5}\\
0 & \frac{1}{18}\sqrt{3}\sqrt{5} & \frac{5}{18}%
\end{bmatrix}
=1-\frac{10}{18}=\frac{4}{9}.$
\end{center}

\noindent The logical entropy is the two-draw probability of drawing a
distinction of $\pi$ so it could also be computed as the sum of all the
distinction probabilities (remembering that a distinction is an ordered pair
of elements in different blocks so the probability of an unordered pair is
doubled). Hence in general we have: $h\left(  \pi\right)  =\sum_{\left(
u_{i},u_{k}\right)  \in\operatorname*{dit}\left(  \pi\right)  }p_{i}p_{k}$, or
in the case at hand:

\begin{center}
$h\left(  \pi\right)  =2p_{a}p_{b}+2p_{a}p_{c}=2\frac{1}{3}\frac{1}{4}%
+2\frac{1}{3}\frac{5}{12}=\frac{1}{6}+\frac{5}{18}=\frac{4}{9}$.
\end{center}

There is then a general theorem \cite{ell:nf4it} showing how logical entropy
measures measurement.

\noindent\textbf{Theorem} (set case of measuring measurement): In the
L\"{u}ders mixture operation $\hat{\rho}\left(  \pi\right)  =\sum_{r\in
g\left(  U\right)  }P_{g^{-1}\left(  r\right)  }\rho\left(  \pi\right)
P_{g^{-1}\left(  r\right)  }$, the increase in logical entropy from $h\left(
\rho\left(  \pi\right)  \right)  $ to $h\left(  \hat{\rho}\left(  \pi\right)
\right)  $ is the sum of the squares of the off-diagonal non-zero entries in
$\rho\left(  \pi\right)  $ that were zeroed in the measurement operation
$\rho\left(  \pi\right)  \rightsquigarrow\hat{\rho}\left(  \pi\right)  $.

In the example, the logical entropy of the post-measurement state is:

\begin{center}
$h\left(  \hat{\rho}\left(  \pi\right)  \right)  =1-\operatorname*{tr}\left[
\hat{\rho}\left(  \pi\right)  ^{2}\right]  =1-\operatorname*{tr}\allowbreak%
\begin{bmatrix}
\frac{1}{9} & 0 & 0\\
0 & \frac{1}{16} & 0\\
0 & 0 & \frac{25}{144}%
\end{bmatrix}
=1-\left(  \frac{1}{9}+\frac{1}{16}+\frac{25}{144}\right)  =1-\frac
{16+9+25}{144}=\frac{94}{144}$.
\end{center}

\noindent The sum of the squares of the non-zero off-diagonal terms
(representing the coherences) of $\rho\left(  \pi\right)  $ that were zeroed
(decohered) in $\hat{\rho}\left(  \pi\right)  $ is:

\begin{center}
$2\left(  \frac{\sqrt{5}}{4\sqrt{3}}\right)  ^{2}=2\frac{5}{48}=\frac{5}{24}$
\end{center}

\noindent and the increase in logical entropy due to the making of
distinctions is:

\begin{center}
$h\left(  \hat{\rho}\left(  \pi\right)  \right)  -h\left(  \rho\left(
\pi\right)  \right)  =\frac{94}{144}-\frac{4}{9}=\frac{94}{144}-\frac{64}%
{144}=\frac{30}{144}=\frac{5}{24}$.$\checkmark$
\end{center}

And the quantum case is \textit{mutatis mutandis}.

\noindent\textbf{Theorem} (quantum case of measuring measurement): In the
L\"{u}ders mixture operation $\hat{\rho}=\sum_{r\in g\left(  U\right)
}P_{V_{r}}\rho P_{V_{r}}$, the increase in quantum logical entropy from
$h\left(  \rho\right)  $ to $h\left(  \hat{\rho}\right)  $ is the sum of the
absolute squares of the off-diagonal entries in $\rho$ that were zeroed in the
measurement operation $\rho\rightsquigarrow\hat{\rho}$.

The dictionary relating the three basic concepts in the math of QM to their
set partitional precursors is given in the following Table 5.

\begin{center}%
\begin{tabular}
[c]{l||l|l|}\cline{2-3}%
Dictionary & Partition math & Hilbert space math\\\hline\hline
\multicolumn{1}{|l||}{Notion of state} & $\rho\left(  \pi\right)  =\sum
_{j=1}^{m}\Pr\left(  B_{j}\right)  \left\vert b_{j}\right\rangle \left\langle
b_{j}\right\vert $ & $\rho=\sum_{i=1}^{n}\lambda_{i}\left\vert u_{i}%
\right\rangle \left\langle u_{i}\right\vert $\\\hline
\multicolumn{1}{|l||}{Notion of observable} & $g=\sum_{r\in g\left(  U\right)
}r\chi_{g^{-1}\left(  r\right)  }:U\rightarrow%
\mathbb{R}
$ & $G=\sum_{r\in g\left(  U\right)  }rP_{V_{r}}$\\\hline
\multicolumn{1}{|l||}{Notion of measurement} & \multicolumn{1}{||c|}{$\hat
{\rho}\left(  \pi\right)  =\sum_{r\in g\left(  U\right)  }P_{g^{-1}\left(
r\right)  }\rho\left(  \pi\right)  P_{g^{-1}\left(  r\right)  }$} & $\hat
{\rho}=\sum_{r\in g\left(  U\right)  }P_{V_{r}}\rho P_{V_{r}}$\\\hline
\end{tabular}

Table 5: Three basic notions: set version and corresponding Hilbert space version
\end{center}

\section{Other aspects of QM mathematics}

\subsection{Commuting, non-commuting, and conjugate operators}

We have seen that a Hermitian operator is the QM math version of a real-valued
numerical attribute and that the direct-sum decomposition of the operator's
eigenspaces is the QM math version of the inverse-image partition of the
numerical attribute. Let $F,G:V\rightarrow V$ be two Hermitian operators with
the corresponding DSDs of $\left\{  V_{j}\right\}  _{j\in J}$ and $\left\{
W_{j^{\prime}}\right\}  _{j^{\prime}\in J^{\prime}}$. Since the two DSDs are
the vector space version of partitions, consider the join-like operation
giving the set of non-zero subspaces formed by the intersections $V_{j}\cap
W_{j^{\prime}}$. The additional generality gained over the join of set
partitions is that these subspaces may not span the whole space $V$. Since the
vectors in those intersections are simultaneous eigenvectors of $F$ and $G$,
let $\mathcal{SE}$ be the subspace spanned by the simultaneous eigenvectors of
$F$ and $G$. The commutator $\left[  F,G\right]  =FG-GF:V\rightarrow V$ is a
linear operator on $V$ so it has a kernel $\ker\left[  F,G\right]  $
consisting of the vectors $v$ such that $\left[  F,G\right]  v=\mathbf{0}$.
Then there is a:

\textbf{Theorem}: $\mathcal{SE}=\ker\left[  F,G\right]  $. \cite[Proposition
1]{ell:ftm}

Since commutativity is defined as $\ker\left[  F,G\right]  =V$, we have the
following definitions in terms of the vector space partitions or DSDs:

\begin{itemize}
\item $F$ and $G$ are \textit{commuting} if $SE=V$;

\item $F$ and $G$ are \textit{incompatible} if $SE\neq V$;

\item $F$ and $G$ are \textit{conjugate} if $SE=0$ (zero space).
\end{itemize}

Since the join-like operation on DSDs yields a set of subspaces that do not
necessarily span the whole space, that operation is only the \textit{join} of
DSDs in the commuting case--or as Hermann Weyl put it: "Thus combination of
two gratings presupposes commutability...". \cite[p. 257]{weyl:phil}

The set version of \textit{compatible} partitions for the join operation is
simply being defined on the same set. Hence our thesis gives a complete
parallelism between compatible partitions and commuting operators.

\textbf{Set math}: A set of compatible partitions $\pi,\sigma,...\gamma$
defined by $f,g,...h:U\rightarrow%
\mathbb{R}
$ is said to be \textit{complete}, i.e., a Complete Set of Compatible
Attributes or CSCA,\textit{ }if their join is the partition whose blocks are
of cardinality one (i.e., $\mathbf{1}_{U}$). Then the elements $u\in U$ are
uniquely characterized by the ordered set of values $(f\left(  u\right)
,g\left(  u\right)  ,...,h\left(  u\right)  $.

\textbf{QM math}: A set of commuting observables $F,G,...,H$ is said to be
\textit{complete}, i.e., a Complete Set of Commuting Observables or CSCO
\cite{dirac:principles}, if the join of their eigenspace DSDs is the DSD whose
subspaces are of dimension one. Then the simultaneous eigenvectors of the
operators are unique characterized by the ordered set of their eigenvalues.

\subsection{Feynman's treatment of measurement}

The partitional approach to understanding the math of QM shows that the key
organizing concepts are indistinction versus distinction, indefiniteness
versus definiteness, or indistinguishability versus distinguishability. When a
particle in a superposition state undergoes an interaction, what characterizes
whether it is a measurement or not? As early as 1951 \cite{feynman:1951},
Richard Feynman gave the analysis of "measurement or not" in terms of distinguishability.

\begin{quotation}
\noindent If you could, \textit{in principle}, distinguish the alternative
\textit{final} states (even though you do not bother to do so), the total,
final probability is obtained by calculating the \textit{probability} for each
state (not the amplitude) and then adding them together. If you
\textit{cannot} distinguish the final states \textit{even in principle}, then
the probability amplitudes must be summed before taking the absolute square to
find the actual probability.\cite[p. 3-9]{feynman:vIII}
\end{quotation}

\noindent This analysis has been further explained by another physicist.

\begin{quotation}
\noindent Feynman's approach is based on the contrast between processes that
are \textit{distinguishable} within a given physical context and those that
are \textit{indistinguishable} within that context. A process is
distinguishable if some record of whether or not it has been realized results
from the process in question; if no record results, the process is
indistinguishable from alternative processes leading to the same end result.
\cite[p. 314]{stachel:needqlogic}
\end{quotation}

Feynman gives a number of examples (\cite[\S \ 3-3]{feynman:vIII}; \cite[pp.
17-8]{feyhibbs:styer-ed}) such as a particle scattering off the atoms in a
crystal. If there is no physical record of which atom the particle scattered
off of (i.e., the indistinguishable case), then no measurement took place so
the amplitudes for the superposition state of scattering off the different
atoms are added to compute the amplitude of the particle reaching a certain
final state. But if all the atoms had, say spin up, and scattering off an atom
flipped the spin, then a physical record exists (i.e., the distinguishable
case) so a measurement took place and the probabilities of scattering off the
different atoms are added to compute the probability of reaching a certain
final state.

The same analysis applies to the well-known double-slit experiment where the
distinguishable case is where there are detectors at the slits and the
indistinguishable case is having no detectors at the slits. But the important
thing to notice about Feynman's example is that the measurement is entirely at
the quantum level; it involves no macroscopic apparatus. Hence the Feynman
analysis bypasses the whole tortured literature trying to analyze measurement
in term of the "decoherence" induced by a macroscopic measuring devices (e.g.,
\cite{zurek:decoh}). Of course, the quantum level physical record in the
distinguishable case has to be amplified for humans to record the result but
such macroscopic considerations have no role in quantum \textit{theory}.

The implicit principle in Feynman's analysis of measurement is:

\begin{center}
\textit{If the interaction distinguishes between superposed eigenstates,}

\textit{then a distinction (state reduction) is made.}

The State Reduction Principle
\end{center}

The mathematics of the State Reduction Principle can be stated in both the set
case and the QM case.

\textbf{Theorem }(State Reduction Principle--set case). Measurement is
described in the set case by the L\"{u}ders mixture operation $\hat{\rho
}\left(  \pi\right)  =\sum_{r\in g\left(  U\right)  }P_{g^{-1}\left(
r\right)  }\rho\left(  \pi\right)  P_{g^{-1}\left(  r\right)  }$. The State
Reduction Principle then states: if an off-diagonal entry $\rho\left(
\pi\right)  _{ik}\neq0$ (i.e., $u_{i}$ and $u_{k}$ are in a same-block
superposition), then: if $g\left(  u_{i}\right)  \neq g\left(  u_{k}\right)  $
(i.e., the interaction distinguishes $u_{i}$ and $u_{k}$), then $\hat{\rho
}\left(  \pi\right)  _{ik}=0$ (i.e., the `coherence' between $u_{i}$ and
$u_{k}$ is decohered and a distinction is made).

\textbf{Theorem }(State Reduction Principle--QM case). Measurement is
described in QM by the L\"{u}ders mixture operation $\hat{\rho}=\sum_{r\in
g\left(  U\right)  }P_{V_{r}}\rho P_{V_{r}}$ (measuring $\rho$ by $G$). The
State Reduction Principle then states: if an off-diagonal entry $\rho_{ik}%
\neq0$ (i.e., the $G$-eigenvectors $\left\vert u_{i}\right\rangle $ and
$\left\vert u_{k}\right\rangle $ are in a superposition in $\rho$), then: if
$\left\vert u_{i}\right\rangle $ and $\left\vert u_{k}\right\rangle $ have
different $G$-eigenvalues (i.e., the vectors are distinguished by $G$), then
$\hat{\rho}_{ik}=0$ (i.e., the vectors are decohered\footnote{This is not the
Zeh/Zurek "decoherence" \cite{zurek:decoh} but the old-fashioned change from
the coherence of a superposition pure state into a decohered mixture of
states.} and a distinction is made).

If no distinctions were made by the interaction, then no measurement took place.

\subsection{Von Neumann's type I and type II processes}

John von Neumann made his famous distinction between the processes:

\begin{enumerate}
\item Type I process of measurement and state reduction, and

\item Type II process obeying the Schr\"{o}dinger equation.
\end{enumerate}

We have seen that the Type I processes of measurement involves
distinguishability, i.e., the making of distinctions (like which atom the
particle scattered off of), so a natural way to designate the Type II
processes would be ones that do not make distinctions by preserving
distinguishability or indistinguishability. The measure of indistinctness of
two quantum states is their overlap or inner product. For instance, two states
have zero indistinctness (zero inner product) then they are fully distinct
(orthogonal). Hence the natural characterization of a Type II process is one
that preserves inner products, i.e., a unitary transformation.\footnote{The
connection to solutions to the Schr\"{o}dinger equation in Hilbert space math
is provided by Stone's Theorem (\cite{stone:thm}; \cite[p. 114]%
{hughes:structure}).}

The partitional approach highlights the key analytical concepts of
indistinctions versus distinctions and the cognate notions of indefiniteness
versus definiteness or indistinguishability versus distinguishability. Many
people working on quantum foundations seem to ignore those key concepts, and
then the division between the measurement and unitary evolution seems
unfounded, if not "unbelievable."

\begin{quotation}
\noindent\lbrack I]t seems unbelievable that there is a fundamental
distinction between \textquotedblleft measurement\textquotedblright\ and
\textquotedblleft non-measurement\textquotedblright\ processes. Somehow, the
true fundamental theory should treat all processes in a consistent, uniform
fashion. \cite[p. 245]{norsen:fqm}
\end{quotation}

\subsection{Hermann Weyl's imagery for measurement}

An industrial sieve is used to distinguish particles of matter of different
sizes so it might serve as a helpful metaphor for the quantum process of
making distinctions, namely measurement.

\begin{quotation}
\noindent In Einstein's theory of relativity the observer is a man who sets
out in quest of truth armed with a measuring-rod. In quantum theory he sets
out armed with a sieve.\cite[p. 267]{edd:pathways}
\end{quotation}

Hermann Weyl quotes Eddington's passage \cite[p. 255]{weyl:phil} but uses his
own expository notion of a "grating." Weyl in effect uses the Yoga from the
mathematical folklore to develop both the set notion of a grating as an
"aggregate [which] is used in the sense of `set of elements with equivalence
relation.'" \cite[p. 239]{weyl:phil} and the vector space notion of a
direct-sum decomposition. In the set to vector space move of the Yoga, the
"aggregate of $n$ states has to be replaced by an $n$-dimensional Euclidean
vector space" \cite[p. 256]{weyl:phil} ("Euclidean" is an old name for an
inner product space). The notion of a vector space partition or "grating" in
QM is a "splitting of the total vector space into mutually orthogonal
subspaces" so that "each vector $\overrightarrow{x}$ splits into $r$ component
vectors lying in the several subspaces" \cite[p. 256]{weyl:phil}, i.e., a DSD.
After thus referring to a partition and a DSD as a "grating" or "sieve," Weyl
notes that "Measurement means application of a sieve or grating" \cite[p.
259]{weyl:phil}, i.e., the making of distinctions by the join-like process
described by the L\"{u}ders mixture operation.

This imagery of measurement as passing through a sieve or grating is
illustrated in Figure 6.%

\begin{center}
\includegraphics[
height=1.7141in,
width=1.9718in
]%
{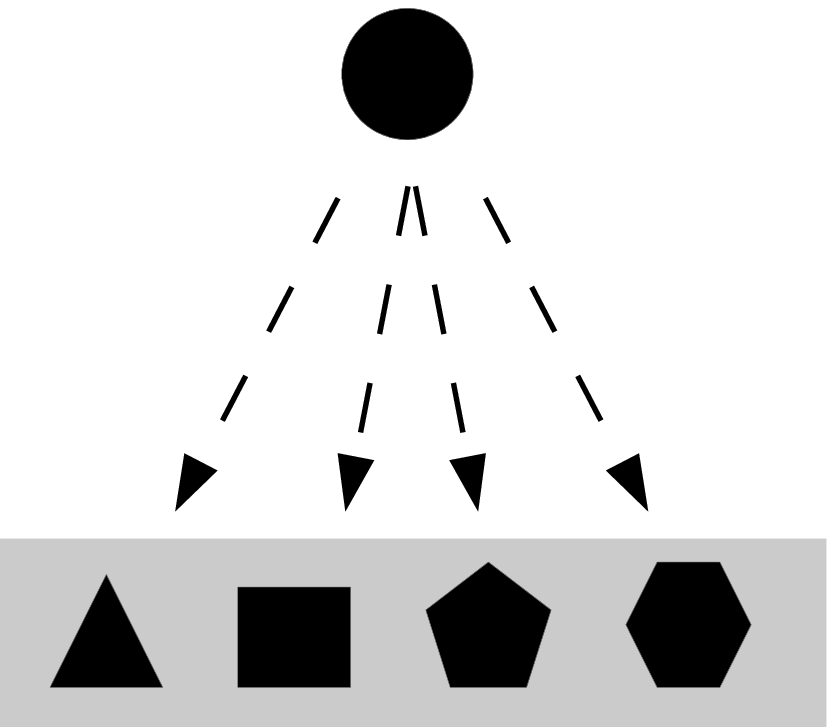}%
\end{center}

\begin{center}
Figure 6: Measurement imaged as an indefinite blob of dough passing through a
grating to get a definite shape
\end{center}

\noindent One should imagine the roundish blob of dough as the superposition
of the definite shapes in the grating or sieve. The interaction between the
superposed blob and the sieve/grating forces a distinction, so a distinction
is made as the blob must pass through one of the definite-shaped holes. In
general, a state reduction (`measurement') from an indefinite superposition to
a more definite state takes place when the particle in the superposition state
undergoes an interaction that distinguishes the superposed states.

\subsection{A skeletal analysis of the double-slit experiment}

Consider the skeletal case of a particle have three possible states
$U=\left\{  a,b,c\right\}  $ which are interpreted as vertical positions in
the setup for the double-slit experiment in Figure 7.%

\begin{center}
\includegraphics[
height=1.2471in,
width=2.9308in
]%
{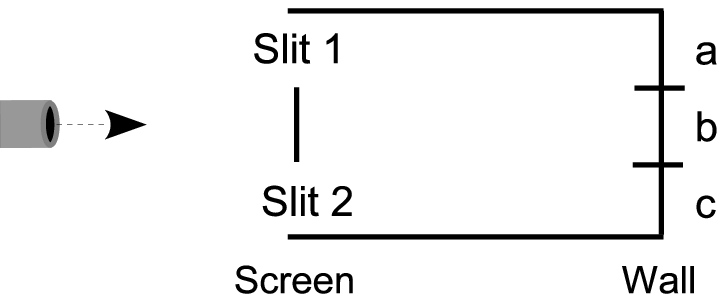}%
\end{center}

\begin{center}
Figure 7:\ Skeletal setup for the double-slit experiment
\end{center}

\noindent In the set level skeletal analysis, we have discarded the scalars
from $%
\mathbb{C}
$ but we are nevertheless left with the scalars $0$ and $1$ which are the
elements of the field $%
\mathbb{Z}
_{2}$. There is the natural correspondence between the zero-one vectors in the
three-dimensional vector space $%
\mathbb{Z}
_{2}^{3}$ (i.e., the column vectors $\left[  1,0,0\right]  ^{t}$ is associated
with $\left\{  a\right\}  $, and so forth) which establishes an isomorphism: $%
\mathbb{Z}
_{2}^{3}\cong\wp\left(  U\right)  $, where the set addition is the symmetric
difference, i.e., for $S,T\in\wp\left(  U\right)  $, $S+T=\left(  S-T\right)
\cup\left(  T-S\right)  $. That mimics the addition mod $2$ in $%
\mathbb{Z}
_{2}^{3}$ since, for instance, $\left\{  a,b\right\}  +\left\{  b,c\right\}
=\left\{  a,c\right\}  $. For our dynamics, we assume a non-singular linear
transformation $\left\{  a\right\}  \rightsquigarrow\left\{  a^{\prime
}\right\}  =\left\{  a,b\right\}  $, $\left\{  b\right\}  \rightsquigarrow
\left\{  b^{\prime}\right\}  =\left\{  a,b,c\right\}  $, and $\left\{
c\right\}  \rightsquigarrow\left\{  c^{\prime}\right\}  =\left\{  b,c\right\}
$ which is non-singular since $\left\{  a^{\prime}\right\}  =\left\{
a,b\right\}  $, $\left\{  b^{\prime}\right\}  =\left\{  a,b,c\right\}  $, and
$\left\{  c^{\prime}\right\}  =\left\{  b,c\right\}  $ also form a basis set
for $\wp\left(  U\right)  $--so we also have a partition lattice $\Pi\left(
U^{\prime}\right)  $ on the basis set $U^{\prime}=\left\{  a^{\prime
},b^{\prime},c^{\prime}\right\}  $.

We are interested in the analysis when the particle arrives at the screen in
the superposition of $\left\vert \text{slit }1\right\rangle +\left\vert
\text{slit }2\right\rangle $, or in skeletal terms $\left\{  a,c\right\}  $.

\textbf{Case 1}: There are detectors at the slits to distinguish between the
two superposed states so the state reduces to the half-half mixture of
$\left\{  a\right\}  $ and $\left\{  c\right\}  $. Then $\left\{  a\right\}  $
evolves by the non-singular dynamics to $\left\{  a,b\right\}  $ which hits
the wall and reduces to $\left\{  a\right\}  $ or $\left\{  b\right\}  $ with
half-half probability. Similar $\left\{  c\right\}  $ evolves to $\left\{
b,c\right\}  $ which hits the wall and reduces to $\left\{  b\right\}  $ or
$\left\{  c\right\}  $ with half-half probability. Since this is the case of
distinctions between the alternative paths to $\left\{  a\right\}  $,
$\left\{  b\right\}  $, or $\left\{  c\right\}  $ we add the probabilities to obtain:

\begin{center}
$\Pr\left(  \left\{  a\right\}  \text{ at wall%
$\vert$%
}\left\{  a,c\right\}  \text{ at screen}\right)  $

$=\Pr\left(  \left\{  a\right\}  \text{ at wall%
$\vert$%
}\left\{  a\right\}  \text{ at screen}\right)  \Pr\left(  \left\{  a\right\}
\text{ at screen%
$\vert$%
}\left\{  a,c\right\}  \text{ at screen}\right)  =\frac{1}{2}\frac{1}{2}%
=\frac{1}{4}$.

$\Pr\left(  \left\{  b\right\}  \text{ at wall%
$\vert$%
}\left\{  a,c\right\}  \text{ at screen}\right)  $

$=\Pr\left(  \left\{  b\right\}  \text{ at wall%
$\vert$%
}\left\{  a\right\}  \text{ at screen}\right)  \Pr\left(  \left\{  a\right\}
\text{ at screen%
$\vert$%
}\left\{  a,c\right\}  \text{ at screen}\right)  $

$+\Pr\left(  \left\{  b\right\}  \text{ at wall%
$\vert$%
}\left\{  c\right\}  \text{ at screen}\right)  \Pr\left(  \left\{  c\right\}
\text{ at screen%
$\vert$%
}\left\{  a,c\right\}  \text{ at screen}\right)  =\frac{1}{2}\frac{1}{2}%
+\frac{1}{2}\frac{1}{2}=\frac{1}{2}$.

$\Pr\left(  \left\{  c\right\}  \text{ at wall%
$\vert$%
}\left\{  a,c\right\}  \text{ at screen}\right)  $

$=\Pr\left(  \left\{  c\right\}  \text{ at wall%
$\vert$%
}\left\{  c\right\}  \text{ at screen}\right)  \Pr\left(  \left\{  c\right\}
\text{ at screen%
$\vert$%
}\left\{  a,c\right\}  \text{ at screen}\right)  =\frac{1}{2}\frac{1}{2}%
=\frac{1}{4}$.
\end{center}

Hence the probability distribution in the Case 1 of measurement at the screen
is given in Figure 8.%

\begin{center}
\includegraphics[
height=1.2471in,
width=2.0911in
]%
{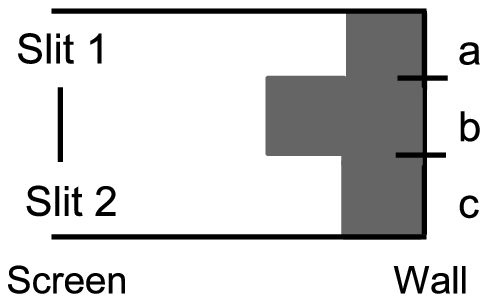}%
\end{center}

\begin{center}
Figure 8: Probabilities at the wall with distinctions at the screen
\end{center}

Case 2: There are no detectors to distinguish between the slits in the
superposition $\left\{  a,c\right\}  $ so it linearly evolves by the dynamics:
$\left\{  a,c\right\}  =\left\{  a\right\}  +\left\{  c\right\}
\rightsquigarrow\left\{  a,b\right\}  +\left\{  b,c\right\}  =\left\{
a,c\right\}  $. Hence the probabilities at the wall are:

\begin{center}
$\Pr\left(  \left\{  a\right\}  \text{ at wall
$\vert$%
}\left\{  a,c\right\}  \text{ at screen}\right)  $

$=\Pr\left(  \left\{  a\right\}  \text{ at wall%
$\vert$%
}\left\{  a,c\right\}  \text{ at wall}\right)  \Pr\left(  \left\{
a,c\right\}  \text{ at wall%
$\vert$%
}\left\{  a,c\right\}  \text{ at screen}\right)  =\frac{1}{2}\times1=\frac
{1}{2}$.

$\Pr\left(  \left\{  b\right\}  \text{ at wall%
$\vert$%
}\left\{  a,c\right\}  \text{ at screen}\right)  $

$=\Pr\left(  \left\{  b\right\}  \text{ at wall%
$\vert$%
}\left\{  a,c\right\}  \text{ at wall}\right)  \Pr\left(  \left\{
a,c\right\}  \text{ at wall%
$\vert$%
}\left\{  a,c\right\}  \text{ at screen}\right)  =0\times1=0$.

$\Pr\left(  \left\{  c\right\}  \text{ at wall
$\vert$%
}\left\{  a,c\right\}  \text{ at screen}\right)  $

$=\Pr\left(  \left\{  c\right\}  \text{ at wall%
$\vert$%
}\left\{  a,c\right\}  \text{ at wall}\right)  \Pr\left(  \left\{
a,c\right\}  \text{ at wall%
$\vert$%
}\left\{  a,c\right\}  \text{ at screen}\right)  =\frac{1}{2}\times1=\frac
{1}{2}$.
\end{center}

Hence the probability distribution in the Case 2 of no distinctions at the
screen is given in Figure 9.%

\begin{center}
\includegraphics[
height=1.2782in,
width=2.1041in
]%
{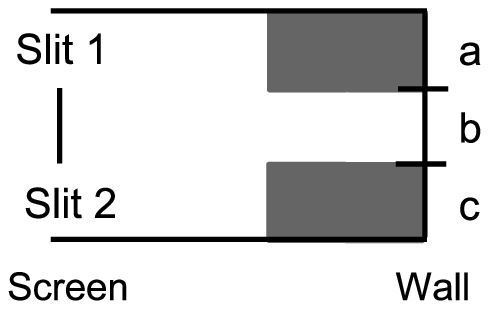}%
\end{center}

\begin{center}
Figure 9: Probabilities at the wall with no distinctions at the screen
\end{center}

\noindent The Case 2 distribution shows the usual probability stripes due to
the interference in the linear evolution of the superposition state $\left\{
a,c\right\}  $, i.e., the destructive interference in the evolved
superposition $\left\{  a,b\right\}  +\left\{  b,c\right\}  =\left\{
a,c\right\}  $.

Our classical intuitions insist on asking: "Which slit did the particle go
through in Case 2?". That question assumes that the evolution of the state
$\left\{  a,c\right\}  $ was at the classical level where the slits were
distinguished. But in Case 2, the slits were not distinguished so the
evolution took place at the lower level in the skeletal lattice of partitions.
In Figure 10, the Case 2 evolution is illustrated as going from the
superposition state $\left\{  a,c\right\}  $ in the partition lattice of
states on $U=\left\{  a,b,c\right\}  $ to the superposition state $\left\{
a^{\prime},c^{\prime}\right\}  $ lattice of states on $U^{\prime}=\left\{
a^{\prime},b^{\prime},c^{\prime}\right\}  $.%

\begin{center}
\includegraphics[
height=1.4935in,
width=4.8153in
]%
{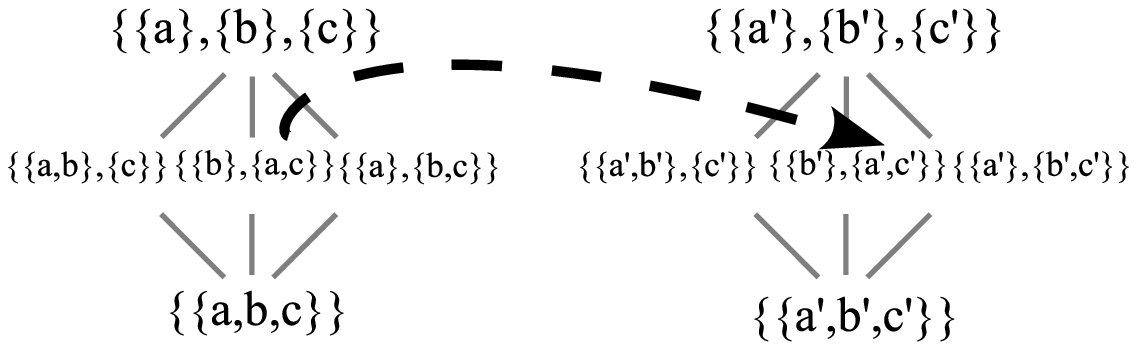}%
\end{center}

\begin{center}
Figure 10: Evolution taking place at a non-classical level of indefiniteness
\end{center}

The important `take-away' is that there are different levels of indefiniteness
(as illustrated in the partition lattice) and evolution can take place at a
non-classical level of indefiniteness so, in that case, there is no matter of
fact of the particle going through one slit or the other at the classical level.

Sometimes metaphors can serve as an aid or crutch to our biologically evolved
intuitions. Consider the "hawks and hounds" of Shakespeare's Sonnet 91. There
is a high fence across a field with two slits or gates. To get from A on one
side of the fence to B on the other side, the hound (like a classical
particle) is limited to horizontal `classical' trajectory on the "flatland"
\cite{abbott:flatland} so it has to go through one gate of another. But the
hawk's "flights and perches" \cite[p. 198]{hawkins:lang-nature} can go from
ground perch A to ground perch B without going through one gate or the other.
We make the unrealistic assumption that a light source above the hawk (like
the sun) will `project' the hawk down to the `classical' definite ground (like
Icarus!). Then the grounded hawk, like the hound, must go through one gate or
the other to get from A to B. But with no light source, then the hawk (like
Hegel's owl of minerva who only flies at night) has an indefinite flight
trajectory and can go from A to B without going through a gate. Our classical
("flatlander") intuitions see only the definite ground-level paths or
trajectories and, in the absence of either projecting light source as in Case
2 above, will insist on asking: "Which gate did the hawk go through?". But
with no detections at the slits in the double-slit experiment, there is no
matter of fact of the particle going through a slit at the classical level
since the evolution is at the non-classical quantum level (illustrated by the
third dimension in our flatlander metaphor) as in Figure 10.

\section{Final remarks}

Our thesis is that the math of QM is the Hilbert space version of the math of
partitions, or, put the other way around, the math of partitions is the
skeletonized version of QM math. There are many other aspects of QM math that
could be investigated such as group representations on sets or on vector
spaces over $%
\mathbb{C}
$ since a group is essentially a `dynamic' algebraic way to define an
equivalence relation or DSD \cite{castellani:symeq} (e.g., the orbit partition
in a set representation or the DSD of irreducible subspaces in the vector
space over $%
\mathbb{C}
$ representation). \cite{ell:ftm} But in this introductory treatment, we have
hopefully analyzed enough aspects of QM math to illustrate our thesis.

Since partitions are the mathematical tool to analyze indistinctions and
distinctions or indefiniteness and definiteness, the thesis shows that the key
QM notion of superposition should be interpreted in terms of (objective)
indefiniteness, and that measurement should be interpreted as an interaction
that makes distinctions so it turns an indefinite state into a state with more
definiteness. This approach to better understanding or interpreting QM works
with the standard von Neumann/Dirac quantum theory. It does not involve any
new physics, unlike the pilot-wave or spontaneous localization theories, or
any many-worldly interpretations of measurement. In that sense, the
partitional approach shows how to develop Shimony's idea of the Literal
Interpretation of the math or "formalism of quantum mechanics" \cite[pp.
6-7]{shim:vienna}. Furthermore, the partitional analysis substantiates the
analysis of Heisenberg, Shimony, and others which describes the quantum world
in terms of potentialities or latencies, where, in both cases, the key
attribute was the \textit{reality} of objective indefiniteness. Hence this way
of understanding or interpreting quantum mechanics might be called the
\textit{Objective Indefiniteness }or\textit{ Literal Interpretation}
(\cite{ell:ftm}, \cite{ell:ijqf}).

\end{document}